\documentclass[11pt]{article}
\usepackage{graphicx}

\newcommand{\BABARPubYear}    {04}

\newcommand{\BABARConfNumber} {17}
\newcommand{\SLACPubNumber} {10586}

\newcommand{\sig}{\ensuremath{\sigma}}

\input pubboard/babarsym

\def\kaons  {\ensuremath{K^{(*)}}\xspace}

\def\pion  {\ensuremath{\pi}\xspace}

\def\pom {\ensuremath{\pm}\xspace}
\def\Chi {\ensuremath{\chi}\xspace}

\def\az    {\ensuremath{A_{0}}}

\def\azd   {\ensuremath{|\az|^{2}}}

\def\thetakstar {\ensuremath{\theta_{\Kstar}}}

\def\cthetakstar{\ensuremath{\cos{\thetakstar}}}
\def\sthetakstar{\ensuremath{\sin{\thetakstar}}}

\newcommand{\cq}[1]{\cos^{2}{#1}}

\newcommand{\text}[1]{\mbox{\rm #1}}
\newcommand{\dd}{\text{d}}

\def\chic  {\ensuremath{\chi_{c}}\xspace}

\long\def\inst#1{\par\nobreak\kern 4pt\nobreak
    {\it #1}\par\vskip 10pt plus 3pt minus 3pt}

\begin{document}
{\pagestyle{empty}

\vspace{-1.3cm}

\begin{flushright}
\babar-CONF-\BABARPubYear/\BABARConfNumber \\
SLAC-PUB-\SLACPubNumber \\
\end{flushright}

\par\vskip 5cm

\begin{center}
\Large \bf Search for \boldmath \B \to \chic\ \kaons \unboldmath Decays
\end{center}
\bigskip

\begin{center}
\large The \babar\ Collaboration\\
\mbox{ }\\
\today
\end{center}
\bigskip \bigskip

\begin{center}
\large \bf Abstract
\end{center}
 We report on the search for  the factorization suppressed decays 
 \B \to \chiczero \kaons and 
 \B \to \chictwo \kaons, with \chiczero and \chictwo decaying into \jpsi \g.
 We use a sample of 124 million \BB events collected with the \babar\
 detector at the \pep2\ storage ring at the Stanford Linear Accelerator
 Center. No significant signal is found and upper bounds for the branching fractions
 are obtained. All results are preliminary.
\vfill
\begin{center}

Submitted to the 32$^{\rm nd}$ International Conference on High-Energy Physics, ICHEP 04,\\
16 August---22 August 2004, Beijing, China

\end{center}

\vspace{1.0cm}
\begin{center}
{\em Stanford Linear Accelerator Center, Stanford University, 
Stanford, CA 94309} \\ \vspace{0.1cm}\hrule\vspace{0.1cm}
Work supported in part by Department of Energy contract DE-AC03-76SF00515.
\end{center}

\newpage
} 

\begin{center}
\small

The \babar\ Collaboration,
\bigskip

%
B.~Aubert,
R.~Barate,
D.~Boutigny,
F.~Couderc,
J.-M.~Gaillard,
A.~Hicheur,
Y.~Karyotakis,
J.~P.~Lees,
V.~Tisserand,
A.~Zghiche
\inst{Laboratoire de Physique des Particules, F-74941 Annecy-le-Vieux, France }
% ...

\end{center}\newpage

\section{INTRODUCTION}
\label{sec:Introduction}

Hadronic decays of heavy mesons are not precisely described, despite the
electroweak nature of the quark decay, because the initial and final states
consist  of mesons, not of quarks.
The factorization scheme allows one to make some predictions  though.
Factorization assumes that a weak decay matrix element can be described as the product of 
two independent hadronic currents.
Under the  factorization hypothesis, 
\B \to $c$  $K^{(*)}$ decays are allowed when $c$ = \jpsi, \psitwos or \chicone, but suppressed when $c$ = 
\chiczero or \chictwo  \cite{Suzuki:2002sq}.
In lowest order heavy quark effective theory, there is no $J\ge2$
operator to create the tensor \chictwo from the vacuum.
The decay rate to  \chiczero is zero due to charge conjugation invariance \cite{hcGudrun}.

Belle has recently \cite{BelleChic0} observed \B \to \chiczero \Kp decays, with 
\chiczero \to \pip\pim or \Kp\Km, with a branching fraction surprisingly large based on 
the expectation from factorization and measurements of the \chicone branching fraction.
BaBar has confirmed the observation  \cite{Silvano} with a branching fraction somewhat lower than,
but compatible with, that measured by Belle.

In this document we attempt the detection of \B \to  $\Chi_{c,i}$ \kaons, $i=0,2$,
 using the radiative  \chic\ \to \jpsi \g decays.

\section{THE \babar\ DETECTOR AND DATASET}
\label{sec:babar}
The data used in this analysis were collected with the \babar\ detector
at the \pep2\ storage ring.
They  represent an integrated luminosity of 112.4 \invfb of data taken at the \FourS resonance. 

The \babar\ detector is described elsewhere~\cite{ref:babar}.
Charged particles are detected with a five-layer, double-sided
silicon vertex tracker (SVT) and a 40-layer drift chamber (DCH) with a
helium-based gas mixture, placed in a 1.5-T solenoidal field produced
by a superconducting magnet. The  charged-particle momentum resolution 
is approximately $(\delta p_T/p_T)^2 = (0.0013 \, p_T)^2 +
(0.0045)^2$, where $p_T$ is the transverse momentum   in \gevc. 
The SVT, with a typical
single-hit resolution of 10\mum,  measures the impact
parameters of charged-particle tracks in both the plane transverse to
the beam direction and along the beam.
Charged-particle types are identified from the ionization energy loss
(\dedx) measured in the DCH and SVT, and from the Cherenkov radiation
detected in a ring-imaging Cherenkov device (DIRC). Photons are
identified by a CsI(Tl) electromagnetic calorimeter (EMC) with an
energy resolution $\sigma(E)/E = 0.023\cdot(E/\gev)^{-1/4}\oplus
0.019$. 
Muons are identified in the instrumented flux return (IFR), composed
of resistive plate chambers and layers of iron, which return
the magnetic flux of the solenoid.

\section{ANALYSIS METHOD}
\label{sec:Analysis}

The channels considered here are \B \to \chic\ \kaons with 
\chic\ \to \jpsi \g, \jpsi \to \epem or \mumu;
\kaon is \Kp or \KS (\pip\pim);
 \Kstarz \to \Kp \pim or \KS \piz;
 \Kstarp \to \Kp \piz or  \KS \pip; and \piz \to \g\g.

Multihadron events are selected by demanding a minimum of three
reconstructed charged tracks in the polar-angle range $0.41 <
\theta_{lab} < 2.54~\rad$.  Charged tracks have to be reconstructed in the
DCH and are required to originate at the beamspot: within
1.5\cm in the plane transverse to the beam and 10\cm along the beam.
Events are required to have a primary vertex within 0.5\cm of the
average position of the interaction point in the plane transverse to
the beamline and within 6\cm longitudinally.
Electromagnetic depositions in the calorimeter in the polar-angle
range $0.410 < \theta_{lab} < 2.409~\rad$ that are not associated with
charged tracks, have an energy larger  than 30\mev, and a shower shape
consistent with a photon are taken as photon candidates.
A total energy larger than 4.5\gev in the fiducial regions for
charged tracks plus neutrals is required.
To reduce continuum background, we require the normalized second
Fox-Wolfram moment $R_2$~\cite{FoxWolfram} of the event, calculated
with both charged tracks and neutral clusters, to be less than 0.5.
Charged tracks are required to be in polar-angle  regions for which the
PID efficiency is well-measured.
For electrons, muons, and kaons the acceptable ranges are  0.40 to 2.40, 0.30 to 2.70 and
0.45 to 2.50 rad, respectively.

Event selection was optimized by  maximizing $\epsilon/\sqrt{B}$, where $\epsilon$ is the 
signal efficiency, and $B$ the number of events, after selection was applied.
Candidates for \jpsi mesons are reconstructed in  the \epem and
\mumu decay modes, from a pair of identified leptons that form a good vertex. 
Muon candidates are identified using a neural network selector and are required 
to pass a loose selection for the first muon candidate and a very loose selection 
for the second muon candidate.  
Electron candidates are selected using a likelihood selector and are required to 
pass a loose selection.  
For \jpsi\to\epem decays, electron candidates are combined with photon candidates 
in order to recover some of the energy lost through bremsstrahlung. Photons are 
required to be within 35\mrad in polar angle from the electron track, and to have 
an azimuthal angle intermediate between the initial track direction (estimated by 
subtracting 50\mrad opposite to the bend direction of the reconstructed track) and 
the centroid of the EMC cluster arising from the track.

The lepton-pair invariant mass has to be between 2.95 and 3.18\gevcc for
both lepton flavors. 
The remaining background is mainly due to genuine \jpsi's. 
Candidates for \KS\  consist of oppositely-charged tracks with invariant mass 
between 487 and 510\mevcc and are required to satisfy vertexing conditions. 
The \KS flight length has to be greater than 1 \mm, and its direction
must form an angle with the \KS momentum vector in the plane
perpendicular to the beam line that is less than 0.2\rad.
Charged kaon candidates are identified using a likelihood selector and are required 
to pass a tight selection. 
Photon candidates as defined above are used also for the reconstruction of
$\piz \to \gamma \gamma$ decays. 
A \piz candidate consists of a pair of photon candidates with invariant mass in
the interval 117 -- 152\mevcc and momentum larger than 350 \mevc. 
\Kstar candidates must have a $K\pi$ invariant mass in the range 0.85 -- 0.94\gevcc
around the nominal $\Kstar(892)$ mass \cite{ref:pdg}. 
The \jpsi, \KS, and \piz candidates are constrained to their 
corresponding nominal masses \cite{ref:pdg}.  

The \chic\ candidates are formed from  \jpsi and photon candidates. The photon is required to 
fulfill the same shower shape requirement mentioned above, have  an energy larger than 
 0.15 \gev, and not be part of \piz candidates in the mass range 0.125 -- 0.140 \gevcc.

The  \chic\ and $K^{(*)}$ candidates are combined to form \B candidates. 
Two kinematic variables are used to further remove incorrect \B candidates.
The first is the difference $\Delta E \equiv E^*_B - E^*_{beam}$ between
the \B-candidate energy and the beam energy in the \FourS\ rest
frame. In the absence of experimental effects, reconstructed signal
candidates have $\Delta E = 0$. 
The second is the beam-energy-substituted mass $\mes \equiv (E^{*2}_{beam} -p^{*2}_B)^{1/2}$. 
The energy substituted mass \mes should peak at the \B meson mass 5.279 \gevcc.

For the signal region, $\Delta E$ is required to be between $-35$\mev
and $+20$\mev for channels involving a \piz, and to be between $-20$\mev 
and $+20$\mev otherwise.  If several \B\ candidates are found in an event, the
one having the smallest $|\DeltaE|$ is retained.
\mes is required to be in the 5.274 -- 5.284 \gevcc range.

Studies using simulated samples show that most of the background
events in the \chic\ \Kstar channels are due to non-resonant (NR) \B \to
\chic\ (\jpsi \g)\kaon \pion decays.  Also the observation of the suppressed
\chictwo could be complicated by the presence of the prominent
\chicone peak.  Therefore the search is performed by the observation
of the spectrum of the mass difference $m_{\ellell \g}-m_{\ellell}$.
It was found from Monte Carlo simulation that after the non-resonant events 
were removed from the sample, the expected number of genuine \chic\ \to \jpsi \g decays
was extremely small, 0.2 \pom 0.2 for the \chictwo \Kstarz (\Kp
\pim) and  \chictwo\Kstarp (\Kp \piz) modes, and 
 0.0 \pom 0.2 for all the other \chictwo modes and all the \chiczero modes.

The efficiencies obtained from fits to the mass difference distribution for 
exclusive samples, where one \B decays to the final state under
consideration and the other inclusively, are given in Table
\ref{tab:chic1:eff}.  Note that \chicone exclusive simulated samples
are used in the place of \chictwo, that were not available.

The \chictwo has a negligible natural width and is therefore fitted
with a Gaussian.  The \chiczero has a natural width $\Gamma = 10.1
\mev$ comparable with the measurement resolution $\sigma \approx 10
\mevcc$, and therefore the \chiczero peak is fitted with a Voigtian, 
the convolution of a Breit-Wigner and a Gaussian distribution.

\begin{table}[htbp]
\caption{Efficiencies from fits of the distribution of $m_{\ellell \g}-m_{\ellell}$ for  exclusive samples.
 \label{tab:chic1:eff}}
\begin{center}    
\begin{tabular}{|c|c|c|c|}
\hline
 & \chicone & \chiczero \\
\hline
 \Kstarz (\Kp \pim) &       0.071 \pom     0.001 &   0.066 \pom     0.001 \\
 \Kstarz (\KS \piz) & 0.031 \pom     0.001 &      0.010 \pom     0.000  \\
 \Kstarp (\Kp \piz) &    0.036 \pom     0.001 &  0.031 \pom     0.001 \\
 \Kstarp (\KS \pip) &    0.065 \pom     0.001 &     0.062 \pom     0.001    \\
 \Kp &   0.144 \pom     0.001 &      0.117 \pom     0.002 \\
 \KS &   0.158 \pom     0.001 &     0.126 \pom     0.001  \\
\hline
\end{tabular}
\end{center}
\end{table}

\begin{figure}[htbp]
\begin{center}
\includegraphics[width=0.75\linewidth]{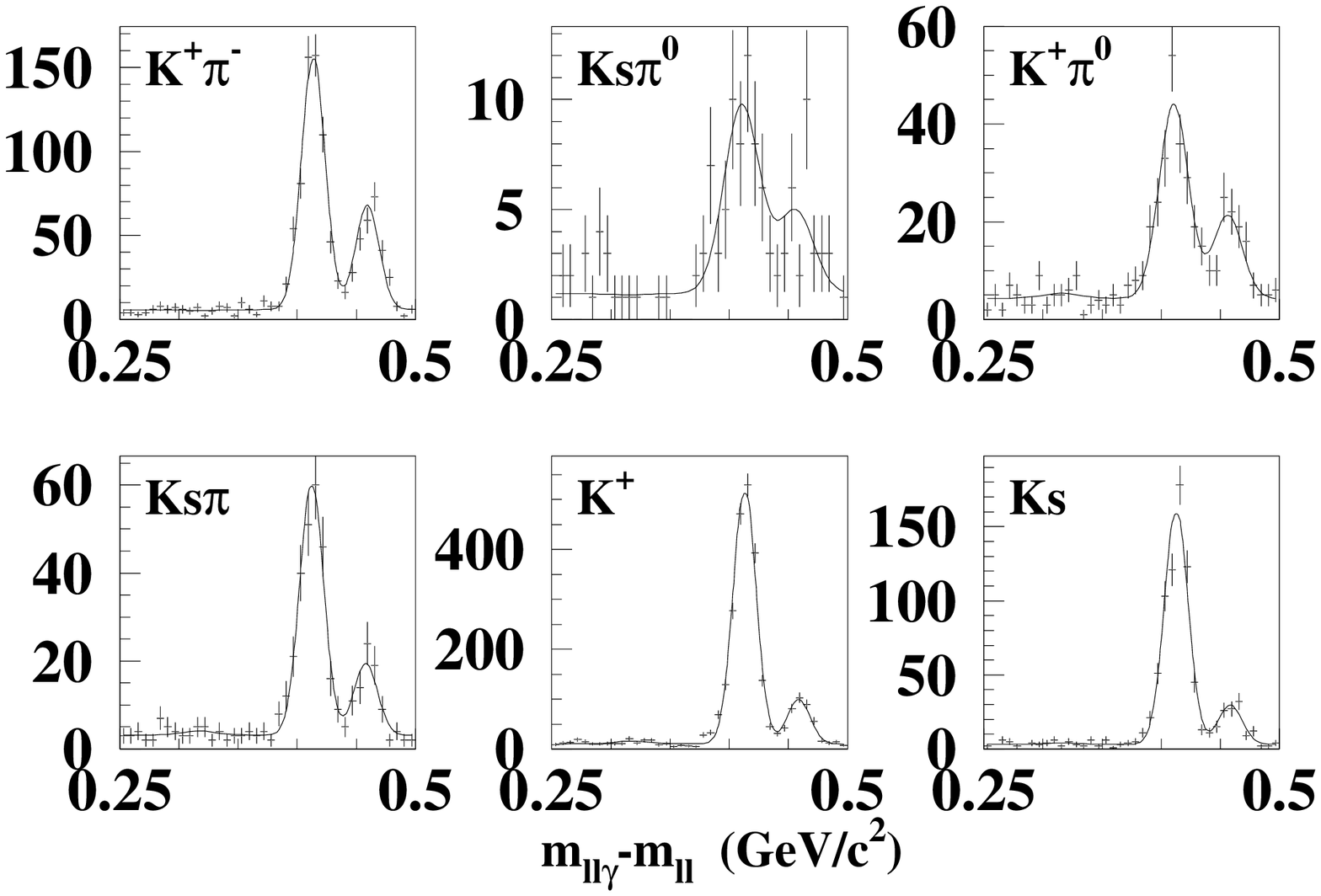}
\caption{Distribution of $m_{\ellell \g}-m_{\ellell}$ (\gevcc) for  generic MC samples.
\label{fig:gene:fit1D}}
\end{center}
\end{figure}

We corrected for the presence of non-resonant decays under the \Kstar peak 
in the following way: the $m_{\ellell \g}-m_{\ellell}$ 
distribution for the events on the plateau; i.e. 
$1.1 < m_{K \pi} < 1.3\gevcc$, is subtracted from the $m_{\ellell \g}-m_{\ellell}$
distribution for the events in the signal region $0.85 < m_{K \pi} <
0.94\gevcc$, after  rescaling by a  factor $r = 0.26 \pom 0.04$, where $r$ is the 
ratio obtained from Monte Carlo simulation of non-resonant events under the 
peak compared to the plateau. 
The branching fractions were then computed from: 

\begin{eqnarray}
  BF = {N \over N_{\B} \times \epsilon \times f}
\end{eqnarray}

where $N$ is the number of events obtained from fitting the $m_{\ellell \g}-m_{\ellell}$
distribution, $N_B$ is the number of \BB events, $\epsilon$ is the selection 
efficiency and $f$ is the secondary branching fraction of the \B daughters. 
Examples of fits to the ``generic'' \BB Monte Carlo (MC) sample, that
contains a simulation of inclusive \FourS \to \BB decays, can be seen
in Fig. \ref{fig:gene:fit1D}.

The free parameters in the fits are: a linear background, the overall mass
difference scale, the resolutions of the gaussian taken to be the same for
the 3 \Chi's and the amplitudes of the peaks. 
The fixed parameters are the natural width of the \chiczero and 2
mass differences, all taken from PDG.

With such fits, it was checked that the non-resonant events were
subtracted correctly, and that the proximity of the \chicone was not
inducing any visible bias on the measurement of the nearby \chictwo.

\section{SYSTEMATIC STUDIES}
\label{sec:Systematics}

This measurement is affected by the following set of systematic uncertainties:

\begin{itemize}
\item Overall uncertainty on the number of \B events, 1.1\%.
\item Uncertainy on the secondary branching fractions: from
PDG \cite{ref:pdg} 
(dominated by the relative uncertainty of the branching fraction of
the radiative decay of the \Chi, 11.9 and 8.5\% for the \chiczero and
\chictwo, respectively).

\item Tracking:  1.3\% per track. 
\item \KS:  a 2.5\%  uncertainty. 

\item Neutrals: 2.5\% per ``\Chi'' photon, 5.0\% per \piz.

\item  An overall 3\% uncertainty on particle identification
correction.

\item Selection cuts: 
For each mass peak and for $\Delta E$,
the uncertainty of the MC-to-data shift in central
value and in width are
measured on the well populated \chicone channels and are
 used to vary the selection cuts, by 1 \sig.
The corresponding efficiency variation, estimated on the exclusive sample, is the induced contribution to
the systematics.  The central value and width induced systematics
are estimated independently, and are added quadratically below.

The results for \chicone MC sample and \chiczero MC sample are
quite close to each other; an average value is used for both.

\item 
The ratio of \Bz to \Bp production  in \FourS decays is assumed to be unity.
The related uncertainty is small \cite{Aubert:2004ur}
and is neglected here.

\item
The NR component is probably in an  S-wave K\pion state,
as was observed in the \jpsi \Kstar system  
\cite{Aubert:2001pe}, with 
an unknown relative phase $\phi$  wrt the main \Kstar(892) 
P-wave peak.

It is possible that no signal is found in the channels under
consideration in this section.  Therefore the systematics due to the
unknown relative phase is here estimated with a MC-based method.

The K\pion invariant mass is fitted with an amplitude that is the sum 
of a non-relativistic Breit-Wigner and a real amplitude that 
corresponds to a polynomial (parabolic) distribution for the NR 
(Fig. \ref{fig:fitsphase}).
\begin{equation}
p(m_{K \pi}) = \left| \frac{a}{m_{pdg}-m_{K \pi}-i\Gamma/2}+b(m_{K
     \pi}) e^{i \phi}\right|^2
\label{eq:BreitPlusNR}
\end{equation}
where $a$ and $b$ are real quantities. The slow variation of the
phase of the S wave with $m_{K \pi}$ is neglected here.
\begin{figure}[htbp]
\begin{center}
\includegraphics[width=0.45\linewidth]{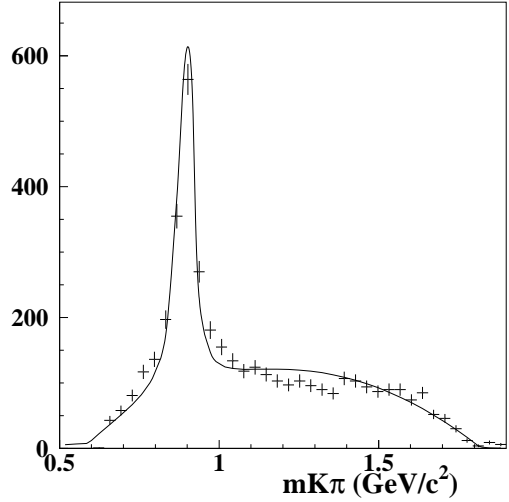}
\caption{
The distribution of $m_{K \pi}$  for the generic MC sample.  
A fit with a coherent sum of a polynomial NR and a 
non-relativistic Breit-Wigner is overlaid.
\label{fig:fitsphase}}
\end{center}
\end{figure}

The free parameters in the fit are the 3 degrees of freedom of the NR
parabola, the magnitude of the signal, and the relative phase $\phi$.
As the high mass plateau is dominated by the NR contribution, no
attempt is made to subtract the few combinatorial events.
The fact that $\phi$ is unknown is dealt with by randomly generating
samples of events distributed as above for each value of $\phi$, and
applying the NR subtraction as described above.  The number of events $N$
thus measured is normalized to that generated with the value $\phi_0$ of $\phi$
obtained in the fit, and the ratio $R=N(\phi)/N(\phi_0)$ is plotted as a function of
$\phi$ in Fig. \ref{varPhaseNR}.
\begin{figure}[htbp]
\begin{center}
\includegraphics[width=0.45\linewidth]{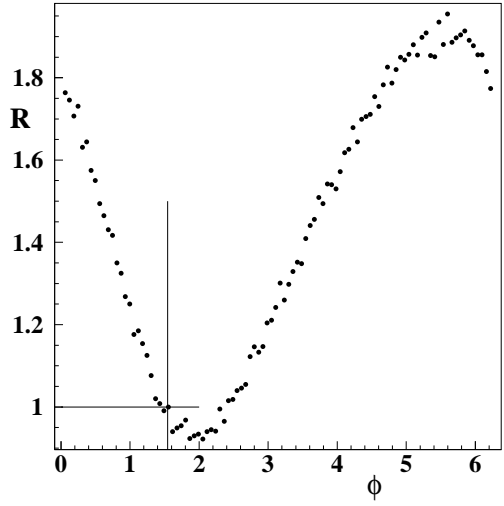}
\caption{Number of events measured, after NR subtraction, 
in Toy MC samples, as a function of    $\phi$, normalized to the number obtained with the phase fitted on the generic sample (shown by the vertical line).
\label{varPhaseNR}}
\end{center}
\end{figure}
The medium value is 1.44 with maximal relative extention \pom 35\%,
giving an RMS relative uncertainty of \pom 20\%.

\item In the case of \chictwo, the efficiency depends on the intensity fractions to various
polarization states, due to the variation of the detection efficiency 
with the angles describing the decay.
The efficiency is mainly sensitive to the \Kstar decay helicity angle,
(Fig. \ref{fig:angle:eff}) due to soft pions for small values of
$\theta_{K*}$.

\begin{figure}[htbp]
\begin{center}
\includegraphics[width=0.65\linewidth]{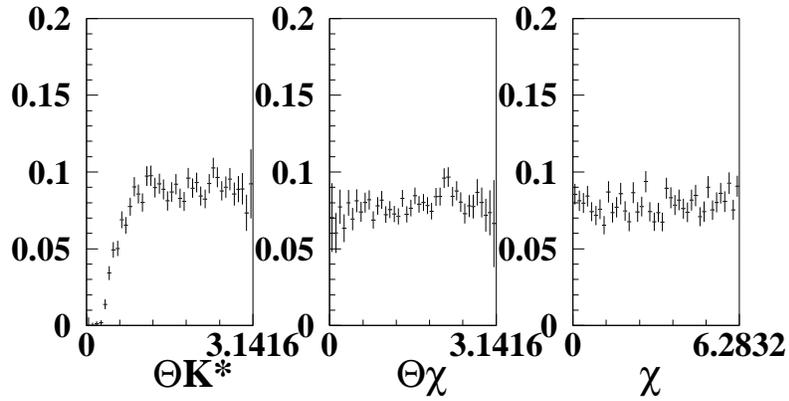}
\caption{Signal efficiency as a function of the helicity angles
 for \Kp \pim channel.
$\chi$ is the angle between the decay planes of the \chic\ and of the \Kstar.
\label{fig:angle:eff}}
\end{center}
\end{figure}

The selection efficiency therefore depends, to 1st order, on the
polarization of the \Kstar population, throught the angular distribution:
\begin{equation}
\frac{1}{\Gamma} 
\frac{\dd\Gamma}{\dd \cthetakstar}=
\frac{3}{4} \left[ (1-\cq{\thetakstar}) + \azd (3\cq{\thetakstar}-1) \right],
\label{1Dcthetakstar}
\end{equation}

where  \azd describes the fraction of longitudinal \Kstar polarization.
The efficiency is:
\begin{equation}
\langle
\varepsilon
\rangle =
\int 
\frac{1}{\Gamma} 
\frac{\dd\Gamma}{\dd \cthetakstar}
\varepsilon (\thetakstar)
\dd \cthetakstar 
 =  a+ \azd b, \nonumber
\label{1DEffcthetakstar}
\end{equation}

where
\begin{equation}
a=
\frac{3}{4}
\int 
 (1-\cq{\thetakstar})
\varepsilon (\thetakstar)\sthetakstar
\dd \thetakstar,
\end{equation}

\begin{equation}
b=
\frac{3}{4}
\int  (3\cq{\thetakstar}-1)
\varepsilon (\thetakstar)\sthetakstar
\dd \thetakstar.
\end{equation}

The values of $a$ and $b$ are obtained from the two above equations and 
from the parametrisation $\varepsilon(\thetakstar)$ extracted from 
Figure \ref{fig:angle:eff} and are shown in Table
\ref{tab:coefs:efficiency:angle}.  In the case no signal is observed,
the polarization is unknown, and we estimate the efficiency as 
$ (a+ 0.5 b) \pm (|b|/\sqrt{12})$.

\begin{table}[htbp]
\caption{Coefficients for the calculation of amplitude dependent
average efficiency for the \chictwo \Kstar channels (\%).
\label{tab:coefs:efficiency:angle}}
\begin{center}       
\begin{tabular}{|c|cccc|}
\hline 
           &   $a$ &   $b$ &  Efficiency    & Fract. uncert. \\ 
\hline 
 \Kstarz (\Kp \pim)  &      8.68 &  -1.40 &   7.98\pom 0.40 &  5.1 \\
 \Kstarz (\KS \piz) &      4.25 &  -1.66 &   3.43\pom 0.48 &  14.0 \\
 \Kstarp (\Kp \piz)  &      5.05 &  -1.79 &   4.16\pom 0.52 &  12.4 \\
 \Kstarp (\KS \pip) &     7.83 &  -1.84 &   6.92\pom 0.53 &  7.7 \\
\hline 
\end{tabular}
\end{center}
\end{table}

As usual the effect is stronger on channels having a \piz in the
final state, as the larger background results in harsher cuts during the
optimization process.

\end{itemize}

A summary of the multiplicative contributions to the systematics can
be found in Table \ref{tab:SumSyst}.

\begin{table}[htbp]
\caption{Systematics: summary of the multiplicative contributions:
relative uncertainties
(\%).
\label{tab:SumSyst}}
\begin{center}       
\begin{tabular}{|c|cccccc|}
\hline 
& \Kstarz (\Kp \pim) &  (\KS \piz) &   (\Kp \piz) &  (\KS \pip) &  \Kp &   \KS \\
\hline 
B counting & 1.1 & 1.1 & 1.1 & 1.1 & 1.1 & 1.1 \\
Tracking   & 5.2 & 2.6 & 3.9 & 3.9 & 3.9 & 2.6 \\
\KS        & --  & 2.5 & -- & 2.5 & -- & 2.5 \\
Neutrals   & 2.5 & 7.5 & 7.5 & 2.5 & 2.5 & 2.5 \\
Particle identification        & 3.0 & 3.0 & 3.0 & 3.0 & 3.0 & 3.0 \\
Cut variation $\langle\rangle$
           & 1.0 & 1.5 & 1.3 & 0.8 & 0.6 & 0.5 \\
Cut variation width
           & 7.6 & 13. & 11.5 & 8.2 & 6.5 & 6.3 \\
MC stat &
 1.4 & 2.9 & 1.7 & 1.8 & 1.3 & 1.3 \\
phase &  20.0 &  20.0 &  20.0 &  20.0 & -- &  -- \\
\chiczero  Sec. BF & 11.9   & 11.9  & 11.9  & 11.9  & 11.9  & 11.9 \\
\hline 
Total for \chiczero & 
      25.4   &   28.3   &   27.6  &    25.5    &  14.8   &   14.6 \\

\hline 
\chictwo  Sec. BF & 8.5  & 8.5  & 8.5  & 8.5  & 8.5 & 8.5 \\ 
Polar & 5.1 & 14.0 & 12.4 & 7.7 & -- & -- \\
\hline 
Total for \chictwo &  
      24.5   &   30.5  &    29.1  &    25.3   &   12.2  &    12.0 \\

 \hline 
\end{tabular}
\end{center}
\end{table}

In addition to these multiplicative contributions comes a contribution
from NR background subtraction.
The contribution of the uncertainty of $r$ is given in Table
\ref{tab:SystNR}.

\begin{table}[htbp]
\caption{Systematics on the measured BF's on the generic samples due to NR subtraction.
 (in units of $10^{-4}$).
\label{tab:SystNR}}
\begin{center} 
\begin{tabular}{|c|cc|}
\hline 
 & \chictwo & \chiczero \\
\hline 
 (\Kp \pim) &  0.0& 0.5 \\
 (\KS \piz) &  0.2& 1.3 \\
 (\Kp \piz) &  0.1& 1.4 \\
 (\KS \pip) &  0.1& 1.7 \\
\hline 
\end{tabular}
\end{center}
\end{table}

\section{PHYSICS RESULTS}
\label{sec:Physics}

\begin{figure}[htbp]
\begin{center}
\includegraphics[width=0.75\linewidth]{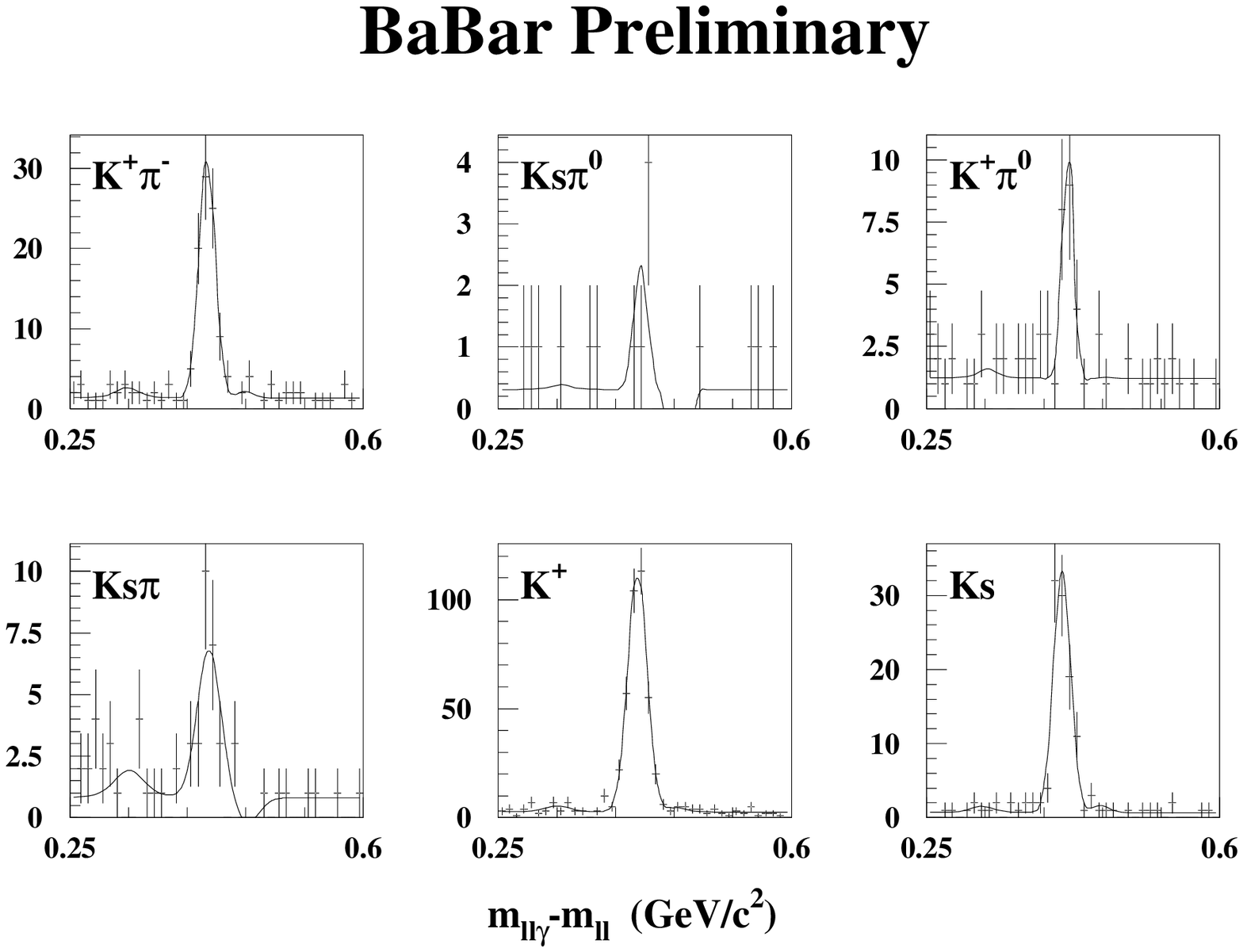}
\includegraphics[width=0.75\linewidth]{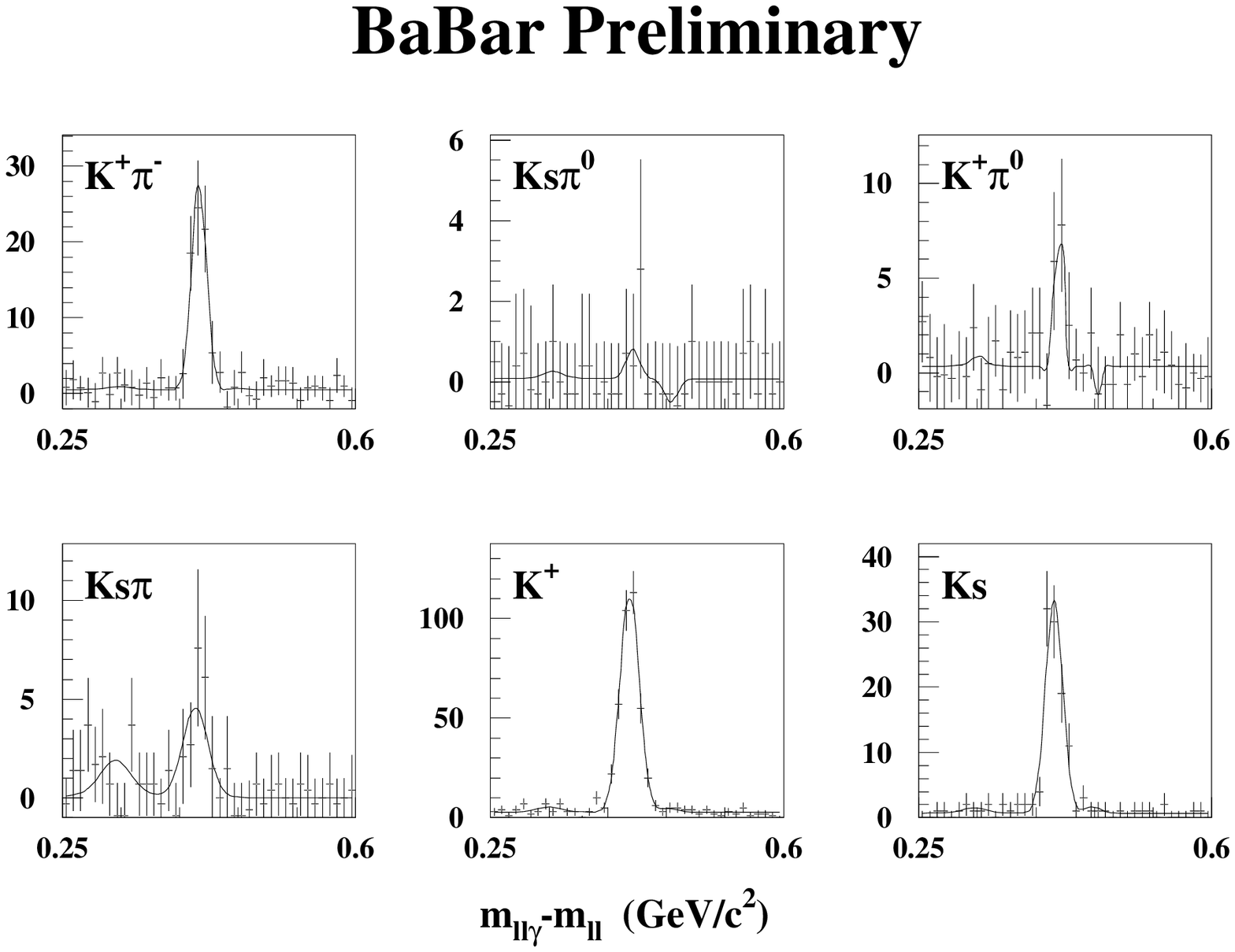}
\caption{ Distribution of $m_{\ellell \g}-m_{\ellell}$ for data.
Top: raw data; Bottom: NR subtracted.
\label{fig:data}}
\end{center}
\end{figure}

Fits on the data clearly show the presence of the factorization allowed 
\chicone, but no signal, within uncertainty, for the factorization suppressed 
\chiczero and \chictwo (Table \ref{tab:fit1d:data}, Fig. \ref{fig:data}).

\begin{table}[htbp]
\caption{Number of events from fits of the distribution of $m_{\ellell \g}-m_{\ellell}$ for the data.
(NR subtracted).
\label{tab:fit1d:data}}
\begin{center} 
    \mbox{ \small
\begin{tabular}{|c|cc|}
\hline 
 & \chictwo & \chiczero \\
\hline 
 \Kstarz (\Kp \pim) &  0.3 \pom 1.1 &  2.1 \pom 2.5 \\
 \Kstarz (\KS \piz) &  -1.7 \pom 2.0 &  1.0 \pom 0.9 \\
 \Kstarp (\Kp \piz) &  -1.8 \pom 0.6 &  0.1 \pom 2.9 \\
 \Kstarp (\KS \pip) &  -0.2 \pom 1.2 &  12.3 \pom 3.7 \\
 \Kp &  6.4 \pom 4.8 &  15.1 \pom 7.6 \\
 \KS &  2.8 \pom 2.6 &  4.5 \pom 4.0 \\
\hline 
\end{tabular}
}
\end{center}
\end{table}

Non-resonant events subtraction has been applied. The phase-related systematics 
estimated on the MC sample are used for the data.

Combining the measurements of the \Kstar sub-modes,  and under the reasonable approximation
that the multiplicative efficiencies between each \Kstar sub-mode are
fully correlated, we obtain the branching fractions for the suppressed modes listed in Table 
\ref{tab:bf:meas}.

The results for the allowed \chicone are found to be compatible with those of \cite{Philippe},
an analysis optimized to the relevant BF, in contrast with this one.

\begin{table}[htbp]
\caption{Measured Branching fractions 
 (in units of $10^{-4}$).\label{tab:bf:meas}}
\begin{center}
\begin{tabular}{|c|cc|}
\hline 
 & \chictwo & \chiczero \\ 
\hline 
\Kstarz & 
 0.02 \pom  0.07 \pom  0.10 &  3.09 \pom  3.35 \pom  1.27 \\
\Kstarp & 
 -0.37 \pom  0.15 \pom  0.20 & 27.0 \pom 11.2 \pom  9.0 \\
\Kp &
 0.15 \pom  0.11 \pom  0.12 &  7.49 \pom  3.76 \pom  1.09 \\
\Kz &
 0.18 \pom  0.16 \pom  0.12 &  5.96 \pom  5.39 \pom  0.88 \\
\hline 
\end{tabular}
\end{center}
\end{table}

Upper bounds on the BF's, at 90\% confidence level (CL) are obtained
using a simulation, assuming gaussian statistics for the statistical
uncertainties and taking into account the systematic uncertainties (Table
\ref{tab:bf:upper}).
It has been assumed that the BF can only be positive.

\begin{table}[htbp]
\caption{Branching fractions: upper bounds at 90\% CL.
 (in units of $10^{-4}$).
 \label{tab:bf:upper}}
\begin{center}
\begin{tabular}{|c|cc|}
\hline 
 & \chictwo & \chiczero \\
\hline 
 \Kstarz & 0.22 & 8. \\
\Kstarp & 0.14 & 45. \\
\Kp & 0.36 & 12. \\
\Kz & 0.44 & 13. \\
\hline 
\end{tabular}
\end{center}
\end{table}

\section{SUMMARY}
\label{sec:Summary}

The  upper limits obtained for decays to \chictwo are more than one order of 
magnitude lower than the branching fractions of the factorization
allowed decays and of the already observed \B \to \chiczero \Kp
decays. All results are preliminary.

\section{ACKNOWLEDGMENTS}
\label{sec:Acknowledgments}

We are grateful for the 
extraordinary contributions of our \pep2\ colleagues in
achieving the excellent luminosity and machine conditions
that have made this work possible.
The success of this project also relies critically on the 
expertise and dedication of the computing organizations that 
support \babar.
The collaborating institutions wish to thank 
SLAC for its support and the kind hospitality extended to them. 
This work is supported by the
US Department of Energy
and National Science Foundation, the
Natural Sciences and Engineering Research Council (Canada),
Institute of High Energy Physics (China), the
Commissariat \`a l'Energie Atomique and
Institut National de Physique Nucl\'eaire et de Physique des Particules
(France), the
Bundesministerium f\"ur Bildung und Forschung and
Deutsche Forschungsgemeinschaft
(Germany), the
Istituto Nazionale di Fisica Nucleare (Italy),
the Foundation for Fundamental Research on Matter (The Netherlands),
the Research Council of Norway, the
Ministry of Science and Technology of the Russian Federation, and the
Particle Physics and Astronomy Research Council (United Kingdom). 
Individuals have received support from 
CONACyT (Mexico),
the A. P. Sloan Foundation, 
the Research Corporation,
and the Alexander von Humboldt Foundation.

\end{document}